\documentclass[12pt]{iopart}

  \expandafter\let\csname equation*\endcsname\relax
  \expandafter\let\csname endequation*\endcsname\relax
\usepackage{amsmath, amssymb}
\usepackage{graphicx}
\usepackage{enumerate}

\begin{document}
\title[Equivalence Principle for Quantum Systems]{Equivalence Principle for Quantum Systems:\\ Dephasing and Phase Shift of Free-Falling Particles}
\author{C. Anastopoulos$^1$ and B. L. Hu$^2$}
\address{$^1$Department of Physics, University of Patras, 26500 Patras, Greece.}
\address{$^2$Maryland Center for Fundamental Physics and Joint Quantum Institute,\\ University of Maryland, College Park, Maryland 20742-4111 U.S.A.}
\ead{anastop@physics.upatras.gr,blhu@umd.edu}

\date{{\small \today}}
\begin{abstract}
We ask the question how the (weak) equivalence principle established in classical gravitational physics should be reformulated and interpreted for massive quantum objects that may also have internal degrees of freedom (dof). This inquiry is necessary because even elementary concepts like a classical trajectory are not well defined in quantum physics -- trajectories originating from quantum histories become viable entities only under stringent decoherence conditions.   From this investigation we posit two logically and operationally distinct statements of the equivalence principle for quantum systems: Version \textbf{A}: The probability distribution of position for a free-falling particle is the same as the probability distribution of a free particle, modulo a {\em mass-independent} shift of its mean. Version \textbf{B}: Any two particles with the same velocity wave-function behave identically in free fall, irrespective of their masses. Both statements apply to all quantum states, including those without a classical correspondence, and also for composite particles with quantum internal dof.

We also investigate the consequences of the interaction between   internal and external dof induced by free fall. For a class of initial states, we find dephasing occurs for the translational dof, namely, the suppression of the off-diagonal terms of the density matrix, in the position basis.  We also find  a gravitational phase shift in the reduced density matrix of the internal dof that  does not depend on the particle's mass.  For classical states, the phase shift has a natural classical interpretation in terms of gravitational red-shift and special relativistic time-dilation.
\end{abstract}
\maketitle

\section{Introduction}

\subsection{Main issues}

Equivalence principle for quantum systems has been a subject of interest amongst theoretical physicists  for a long time, for at least two reasons. First, the equivalence principle (EP) formulated within classical physics plays an important role in the foundation of gravitation. Second, new features introduced by quantum physics could both challenge and enrich the physical contents of the EP. Though often used interchangeably there are subtle differences between the terminology `EP for quantum systems' (EPQS)  and quantum equivalence principle' (QEP): the former assumes the validity of EP for quantum systems, but seeks to explore the new features associated with quantum physics, while QEP may imply a set of new laws applied to quantum systems which are not contained in the classical EPs, or may even be in variance to them. We operate here with the former assumption.

There is one important difference between the classical and quantum contexts of  the EP.  EP in classical physics is expressed in terms of spacetime trajectories. The latter are well defined in classical physics, but not in quantum theory. Trajectories are decohered quantum histories, and thus an emergent concept conditional upon the degree of decoherence. Quantum theory is formulated in terms of state preparation, measurements and probabilities, concepts that are foreign to the classical mind.

The research goal of the EPQS is to explain how the EP can be expressed in terms of such quantum concepts. Of  interest to us in this paper is   to describe quantum particles (both elementary and  composite)  that undergo free-fall in a homogeneous gravitational field and to identify observable consequences. The version of the EP that focuses on test particle motion in a fixed gravitational field---in particular the equivalence of inertial and gravitational mass---is usually referred to as the  Weak Equivalence Principle (WEP).


To avoid extraneous issues arising from more complicated situations we keep our analysis as elementary as possible, which helps to demonstrate that our conclusions are model independent.  Our calculations  involve only non-relativistic quantum mechanics in the weak gravity regime and  measurements carried out in the laboratory reference frame.   However, the context of our calculations is relativistic quantum field theory (QFT). This apparent extra burden is repaid amply by conceptual clarity (especially in the treatment of composite particles) which is very important in dealing with foundational issues like the present one. We presuppose mass-energy equivalence, and we describe the internal degrees of freedom using the language of QFT rather than non-relativistic concepts like the center of mass.


\subsection{Past work}
Prior work on the relation of the EP to quantum theory covers a broad range:  from effects caused by the quantization of gravitational degrees of freedom to the role of quantum internal degrees of freedom in free fall.

Davies and Fang \cite{DaFa82} treated  linearized gravitational perturbations as a massless, spin-two, gauge field coupled to itself and to matter. They  argued that a consistent theory of this type  is impossible unless the gravitational coupling is universal---thus, suggesting the validity of the EP in the quantum domain.

Candelas and Sciama \cite{CanSci83} were interested in the reason why the radiative decay of an atom can be viewed equivalently as a result of the vacuum fluctuations of the EM field or as a result of radiative reactive self force of the electron. Note that the former mechanism is quantum and the latter is classical. They attributed this equivalence to the working of a quantum equivalence principle. 


Alvarez and Mann \cite{CaMa97}  considered possible tests of the EP for physical systems in which quantum-mechanical vacuum energies cannot be neglected. Dalvit and Mazzitelli \cite{DalMaz} calculated graviton-induced corrections to the trajectory of a classical test particle. They showed that the trajectory is no longer a geodesic of the background metric, but a geodesic of the geometry obtained as a solution of the semiclassical Einstein equation that includes backreaction.
Singleton and Wilburn \cite{SinWil}  considered QFT processes like particle creation, in order to  compare the response of an uniformly accelerating detector (Unruh effect, kinematical) to a detector at rest in a Schwarzschild space-time (Hawking effect, gravitational).


As for experimental tests of the EP principle with quantum objects,  there is the famous Colella-Overhauser-Werner experiment \cite{COW}, where the information about free-falling trajectories is encoded in the phase of the wave function and it is retrieved in appropriate interferometric set-ups. There is growing interest in probes of the EP with quantum metrology techniques---for example, Refs. \cite{PCC99, FDHW, SHA}--- because such techniques may allow probing at regimes inaccessible by classical methods.

Closer to the present research is the work of Onofrio and Viola  \cite{OnoVio}. They considered the meaning of the EP in the free fall of quantum particles---in the non-relativistic weak-gravity limit. They found that the quantum mean values are compatible with the classical EP. However, the quantum uncertainties carry a mass-dependence that is incompatible with a naive transposition of the classical EP into the quantum domain. In the same regime, Davies \cite{Dav04} considered tunneling near the classical turning point of a particle in free fall, and showed that tunneling delay could lead to a violation of the classical EP.

The results above appear consistent with the earlier work of Greenberger \cite{Greenb}, who studied a particle bound in an external gravitational potential, and found that  radii and  frequencies depend on the mass of the bound particle, that inertial forces do not look like gravitational forces; and that there are mass-dependent interference effects. He concluded   that the classical statements one can make using the EP  have no direct quantum analog.

Finally, we note the work of  Zych and  Bruckner \cite{ZycBru15, Zych} who emphasized the importance of the internal degrees of freedom of composite particles when formulating quantum versions of the EP. The EP places constraints on the properties of the Hamiltonian that describes the internal degrees of freedom. They argue that this implies that the EP for quantum systems has aspects that are independent of the  classical EP. Orlando et al \cite{OMMP} proposed an experimental test of the EP for composite systems, in which violations of the EP would take the form of forbidden transitions of the internal degrees of freedom.

It is evident from the short summary above that there is no consensus, not only on whether the EP applies to quantum systems, but also on what form it should take. Most of the works admit that the classical statements of the EP cannot be immediately transposed to the quantum context.

In our opinion, this is a consequence of the fundamentally different ontologies that characterize  classical and quantum physics. As mentioned previously, we believe that an EP for quantum systems (if one exists) should be formulated solely in terms of quantum concepts (states, measurements and probabilities) with no reference to classical constructs such as spacetime trajectories.  Under specific conditions where classical physics emerges from quantum physics (such as the existence of quasi-classical domains in sufficiently decohered histories \cite{GeHa}) any newly formulated quantum EP should reproduce the familiar EP stated in terms of classical physics.

\subsection{Our results}

In our present investigation we assume the validity of the classical EP but explore the contents of it when applied to quantum systems.
To this end, we follow a minimalist approach, analyzing a system with transparent enough physics that requires no complex modeling. We consider a non-relativistic free-falling quantum particle in a homogeneous weak gravitational field, taking also into account the internal degrees of freedom. We look for general statements of the EPQS, subject to the following conditions.
\begin{enumerate}
\item It should explain  the mass-dependence in the temporal uncertainties  found in  \cite{OnoVio}.
\item It should apply to all initial states and not only to ones with a classical analogue.
\item It should apply also to composite particles taking into account any effects arising from the internal degrees of freedom.
\end{enumerate}

We found two distinct statements of the EPQS that satisfy the above conditions.
  Both statements involve the comparison between the outcomes of different experiments. For the simple system considered here they are mathematically equivalent. However, they are logically and operationally distinct. They can be tested in different experimental set-ups and they suggest different generalizations.

 The two statements of the EPQS    presented here do not exhaust all possibilities. There is no one-to-one correspondence between classical and quantum concepts, so we should not expect to identify a unique formulation of the EP, at least not in such simple systems. The present forms of the EPQS probably do not apply to more complex set-ups, involving for example,   relativistic effects, strong gravity or  conjectured non-unitary dynamics of gravitational origin. Nonetheless, they provide a stepping stone towards such generalizations. More importantly, they suggest ways of  {\em testing} the EP in presence of distinctly quantum features that are absent in classical physics, i.e.,  non-classical particle states (cat-states) or entanglement between translational and internal degrees of freedom.

 Both statements of the EPQS refer exclusively to measurements carried out by {\em static observers} in the gravitational field. Our statements of the EP cannot be directly applied to statements of the EP pertaining to a comparison of measurements carried out by static and free-falling observers.  In the non-relativistic setting considered here, the comparison of measurement  outcomes for different observers is straightforward. However,  a relativistic generalization would introduce an additional level of complexity, because we would have to take into account the effects of the detector's motion. For example, an accelerated detector may record particles due to the Unruh effect, in which case exchange of quanta between the detector and the field need be included. Even a detector in inertial motion will raise non-trivial issues of transforming outcomes of relativistic position measurement  between different inertial frames.

In this article, we also look for observable consequences of    internal degrees of freedom in free-falling composite particles.
 Assuming only mass-energy equivalence, free fall induces a coupling between the internal and the translational degrees of freedom.    For a particularly chosen class of initial states,  the internal degrees of freedom can lead to a suppression of the off-diagonal terms of the reduced density matrix in the position basis, as it was shown by Pikovsky et al \cite{PikBru15, PikBru17}, modulo possible differences in the interpretation of its meaning. This dephasing is universal in the sense that it applies to all particles and the dephasing time is largely mass-independent, depending only on the properties of the internal degrees of freedom.

Another consequence of the coupling between internal and translational degrees of freedom is a gravitational phase shift in the density matrix of the internal degrees of freedom. While this phase shift is a purely quantum effect, for some initial states it has a natural classical interpretation in terms of gravitational red-shift and special relativistic time-dilation. Similar effects of entanglement between  translational and internal degrees of freedom (dof) show up in the motional decoherence of an atom \cite{Shresta} and in  moving mirrors with internal dof \cite{GBH,SLH}. The difference is that the phase shift derived here is universal, like the dephasing, i.e., it applies to all particles and it is independent of the particle's mass.

This paper is organized as follows: In Sec. 2, we consider a single free-falling elementary particle, where by `elementary' we mean that the particle has no internal structures or their effects can be ignored. We derive two forms of the EPQS and elaborate on their implications. In Secs. 3, we study a free-falling composite particle, i.e., a particle with internal degrees of freedom, such as an atom.  We  show that the EPQS identified in Sec. 2 also apply to composite systems and we recover the dephasing effect of  Ref. \cite{PikBru15}. In Sec. 4, we derive the phase shift in the internal degrees of freedom due to free fall. In Sec. 5, we summarize and discuss our results.

\section{A free-falling particle}
 \subsection{Evolution equations}
Consider a massive particle in free fall in a weak homogenous gravitational field. In this section, we assume that  the particle is elementary in the sense that it has no internal degrees of freedom.
The leading-order terms of the Hamiltonian in the non-relativistic limit are
\begin{eqnarray}
\hat{H}_g = m \hat{1}+ \frac{\hat{p}^2}{2m} + mg \hat{x}, \label{Hamiltonian1}
\end{eqnarray}
where $g$ is the gravitational acceleration. We keep the constant contribution of the rest mass $m$ in the Hamiltonian, because it will be important in the next section where we will consider particles with internal structure.

 In the Appendix, it is shown how Eq. (\ref{Hamiltonian1}) is obtained as the weak-gravity, non-relativistic limit of a free QFT in a static curved spacetime.  The EP  is already invoked in the standard coupling of a quantum field to the background spacetime metric. In Eq. (\ref{Hamiltonian1}) it is assumed that the inertial mass in the kinetic energy term is equal to the gravitational mass in the potential energy term. Inequality between inertial and gravitational mass is possible for QFTs with a gravitational coupling different from that postulated by General Relativity---see an example in the Appendix.

In what follows, we assume the weak equivalence principle holds, namely that the inertial mass $m_i$ equals the gravitational mass $m_g$. (To examine the possibility that $m_i \neq m_g$ one  simply substitutes $m_i$ for $m$ and $m_g g/m_i$ for $g$ in the expressions that follow.)

The Hamiltonian (\ref{Hamiltonian1}) has continuous spectrum over the whole real axis and generalized eigenstates $|E\rangle$,
\begin{eqnarray}
\hat{H}_g |E\rangle = (m+E)|E \rangle.
\end{eqnarray}
We readily evaluate $|E\rangle $ in the momentum representation,
\begin{eqnarray}
\langle p|E\rangle = \frac{1}{\sqrt{2\pi mg}}e^{-\frac{i}{mg} (Ep - \frac{p^3}{6m})}. \label{eigen1}
\end{eqnarray}
The generalized eigenstates (\ref{eigen1}) are normalized so that  $\langle E|E'\rangle = \delta(E-E')$.

We define the propagator
\begin{eqnarray}
G_t^{(g)}(x,x') := \langle x|e^{-i\hat{H}_gt}|x'\rangle = \int dE e^{-i(m+E)t} \langle x|E\rangle \langle E|x'\rangle.
\end{eqnarray}
 Using (\ref{eigen1}) we find
\begin{eqnarray}
G_t^{(g)}(x, x')  = e^{-imgtx - \frac{img^2t^3}{6}} G_t^{(0)}(x + \frac{1}{2}gt^2 ,x'), \label{propag1}
\end{eqnarray}
where
\begin{eqnarray}
G_t^{(0)}(x,x') = \sqrt{\frac{m}{2\pi i t}} e^{ i \frac{m(x-x')^2}{2t}} e^{-imt}
\end{eqnarray}
is the propagator of a free particle.

Let $\psi_0(x)$ be the initial state of the system, and $ \psi_t^{(g)}(x)$ the state at time $t$ evolved with the Hamiltonian $\hat{H}_g$. Eq. (\ref{propag1}) implies that
\begin{eqnarray}
\psi_t^{(g)}(x) = e^{-imgtx - \frac{img^2t^3}{6}} \psi_t^{(0)}(x+ \frac{1}{2}gt^2). \label{psit1}
\end{eqnarray}
Eq. (\ref{psit1}) can be written equivalently as
\begin{eqnarray}
|\psi_t^{(g)}\rangle = e^{i\frac{mg^2t^3}{3}} \hat{V}(-mgt, -\frac{1}{2}gt^2) |\psi_t^{(0)}\rangle, \label{psit2}
\end{eqnarray}
where $\hat{V}(a,b) = e^{ia \hat{x} - i b\hat{p}}$ is the Weyl translation operator.

\subsection{Position  measurements}
Eq. (\ref{psit1}) implies that
\begin{eqnarray}
|\psi_t^{(g)}(x)|^2 = |\psi_t^{(0)}(x+ \frac{1}{2}gt^2)|^2, \label{qep}
\end{eqnarray}
i.e., that the probability distribution of position is the same as that of a free particle with the same initial state, but with a time-dependent shift of the center. The shift {\em does not depend on the particle's mass}.


 Eq. (\ref{qep}) applies to  particles prepared in {\em any initial state}, and not only to particles prepared in a state with a direct classical analogue. In particular, Eq. (\ref{qep})  applies also to cat states, i.e., superpositions of macroscopically distinct configurations.

Eq. (\ref{qep}) describes a probability density defined in the laboratory frame for position at a fixed moment of time. This means that the position of the free-falling particle is determined by a  detector that is static and not in free fall.

The measurement scheme above differs from the classic Galileo-type experiment. In the latter, a particle falls from a fixed height $L$ and the time of arrival at Earth's surface is recorded. Thus, the measured quantity  is the time of arrival, while the location of the particle detector is fixed. Despite the existence of ambiguities in the definition for quantum  time-of-arrival probabilities \cite{ML, ToA},  it is now possible to construct  time-of-arrival probability measures for general Hamiltonians \cite{AnSav}, using a method that can be straightforwardly applied to free-falling particles. However, the method involves more complex techniques of quantum measurement theory. It lies beyond the scope of this paper that aims to find simple characterizations of the EP at the level of elementary quantum mechanics.

\subsection{Saddle point approximation}
Consider  an initial state $\psi_0$ well localized in position and momentum. Let  $\bar{x}_0$ be the mean position   and $\bar{v}_0$ the mean velocity.  In the saddle-point approximation, the integral
\begin{eqnarray}
\psi_t^{(0)}(x) = \int dx' G_t^{(0)}(x,x') \psi_0(x')
\end{eqnarray}
 becomes $ \psi_t^{(0)}(x) = \psi_0(x - \bar{x}_0 - \bar{v}_0 t) e^{-imt}$.
  If the center of the wave-packet is taken as a representative of a classical trajectory,  this expression validates the equivalence principle enacted in classical gravitation theory.

 In the same approximation,   the probability distribution
\begin{eqnarray}
|\psi_t^{(g)}(x)|^2 = |\psi_0(x - \bar{v}_0 t + \frac{1}{2} gt^2) |^2,
\end{eqnarray}
for the position of a free-falling particle is mass-independent.

The latter  conclusion also applies to a large class of non-classical states. Consider, for example, an initial superposition state $\psi_0 = \frac{1}{\sqrt{2}} (\psi_{01} + \psi_{02})$, where the states $\psi_{0i}$ are well localized in position and momentum. Let  $\bar{x}_{0i}$ be the corresponding values for the mean position   and $\bar{v}_{0i}$ for the mean velocity. Then,   the saddle point approximation is a good one, and we obtain
\begin{eqnarray}
 \hspace{-1.5cm} \psi_t^{(g)}(x)  = \frac{1}{\sqrt{2}} \left[  \psi_{01}(x - x_{01} - \bar{v}_{0_1} t + \frac{1}{2} gt^2))  + \psi_{02}(x - x_{02}  - \bar{v}_{02} t + \frac{1}{2} gt^2) \right].
\end{eqnarray}
Again the probability distribution $|\psi_t^{(g)}(x)|^2$ is mass-independent.



 However, the saddle point approximation ignores  the  wave-function dispersion. This is a {\em mass-dependent} process, but this mass dependence has nothing to do with motion in the gravitational field. It is already present in the time evolved  quantum state for a free particle $\psi_t^{(0)}(x)$, and it is not affected by the presence of gravity.

 Consider an initial state with vanishing correlation between momentum and position. Its position uncertainty at time $t$ is
 \begin{eqnarray}
 (\Delta x)^2(t) =  (\Delta x)^2_0  + \frac{(\Delta p)^2_0t^2}{m^2} \geq (\Delta x)^2_0 +  \frac{ t^2}{ 4(\Delta x)^2_0  m^2}, \label{dispers}
 \end{eqnarray}
where $(\Delta x)_0$ and $(\Delta p)_0$ are the initial uncertainties in position and momentum, respectively. Note the uncertainty relation enters in the last step. Eq. (\ref{dispers}) applies both for a free-falling and a free particle.

For $t << m (\Delta x)_0^2$,  $(\Delta x)^2(t) \simeq  (\Delta x)^2_0 $, and the saddle-point approximation holds. However, for $t > m (\Delta x)_0^2$ wave-function dispersion becomes significant and, consequently,  $|\psi_t^{(g)}(x)|^2 $ is {\em not mass-independent}. This mass-dependence of the probabilities due to dispersion has been noticed by \cite{OnoVio} in their study of quantum free-falling particle.

\subsection{Formulation of the equivalence principle}

The analysis above implies that the equivalence principle for quantum systems does {\em not}  take the simple form, that the probability distribution for position for a free-falling particle is mass-independent. Instead, a significantly weaker statement holds.
 \\
\\
{\bf Equivalence principle for quantum systems, Version A:} The probability distribution of position for a free-falling particle is the same as the probability distribution of a free particle, modulo a {\em mass-independent} shift of its mean.
\\ \\
This statement of the equivalence has a direct operational implementation.  It compares the outcome of two experiments that are carried out with the same type of particles and with the same state preparation. In one experiment, the particles move under the influence of an external homogeneous gravitational field and in the other they evolve freely. Both experiments should measure the same position moments $\langle x^n\rangle$ at the same time $t$ except for $n =1$. In that case, the moments differ by a time-dependent term $\frac{1}{2} gt^2$. This term should be the same for different pairs of experiments,  involving different state preparations and even different particles.

In spite of its simplicity, version A of the EP has profound implications. First, it provides a well defined operational procedure for testing the equivalence principle in new regimes, where it has not yet been tested. This is due to the fact that Eq. (\ref{qep}) applies for {\em any} initial state, and not only ones with a classical analogue. For example, it could apply to cat states in a Galileo experiment, where the initial state of a particle is a superposition of two states localized at macroscopic separation $\ell$. This statement of the EP  also implies that wave function dispersion is the same for a free and a free-falling particle of the same type, which is a non-trivial prediction that is in principle testable. Version A is  compatible with the results of Ref. \cite{OPM17}, where it was shown that two-slit interference patterns fall like particles in a homogeneous gravitational field.

Second, this form of the EP is also valid for composite particles. It remains unaffected by the coupling between internal and translational degrees of freedom that is induced by free-fall. We will demonstrate this in the following section.

Third, the EP could turn out to be   important in discussions of gravitational decoherence. The idea of a fundamental decoherence mechanism of gravitational origin has a long history \cite{Karol, Diosi, Penrose} and has been actively pursued in recent years. There are many different models for gravitational decoherence that lead to different predictions. Most models are applied to the decoherence of non-classical states for free particles. However, if the origin of the fundamental decoherence is gravitational, it is natural to inquire whether it conforms to the symmetries and properties of the current theory of gravity, including the EP. Thus, the quantum statements  of the  EP could provide important theoretical constraints to models of gravitational decoherence.

Version A of the EPQS refers to  measurements by a static detector in a weak gravitational field. Both notions (static detector, weak field) can straightforwardly be transferred to the relativistic context. In particular, for static detectors in a weak field particle creation effects are negligible, so we can still use a particle description. Any obstacle to a relativistic generalization of the EPQS would arise from the ambiguities in the relativistic description of position measurements---see, for example, Ref. \cite{Bush}.

\subsection{Alternative formulation of the equivalence principle}
We can provide a different formulation of the equivalence principle by examining free fall in the Wigner picture. For any density matrix $\hat{\rho}$, we define the Wigner function
\begin{eqnarray}
W(x, p) = \int \frac{d\xi}{2\pi}e^{-ip\xi} \langle x+\frac{1}{2}\xi|\hat{\rho}|x -\frac{1}{2}\xi\rangle
\end{eqnarray}
as a quasi-probability density on the classical phase space $\Gamma$ spanned by position $x$ and momentum $p$ .

By Eq. (\ref{psit2}), we readily evaluate the Wigner function $W_t^{(g)}(x,p)$ at time $t$ for a particle of mass $m$ freely falling in a gravitational field $g$,
\begin{eqnarray}
W_t^{(g)}(x,p) = W_0( x - \frac{p}{m}t + \frac{1}{2}gt^2, p+ mg t), \label{wigev}
\end{eqnarray}
where $W_0$ is the Wigner function at time $t = 0$. In this case, the time evolution of the Wigner function  is identical to the time evolution of a classical probability distribution according to the Liouville equation.

It is evident that the   mass-dependence in the Wigner function originates from its momentum dependence.  If we change variables to
  $(x, v)$ where $v = p/m$ is the velocity, the mass-dependence disappears. To this end we define the velocity Wigner function
  \begin{eqnarray}
  \bar{W}(x,v) = \frac{1}{m} W(x, mv), \label{velwig}
  \end{eqnarray}
which allows us to rewrite Eq. (\ref{wigev}) as
\begin{eqnarray}
\bar{W}_t^{(g)}(x,v) = \bar{W}_0( x - vt + \frac{1}{2}gt^2, v + g t). \label{wigev2}
\end{eqnarray}
Two particles of different masses $m_1$ and $m_2$,  but with the same initial velocity Wigner function {\em behave exactly the same} in free fall. Equality of the velocity Wigner function implies that  the state vectors $|\psi_1\rangle $ and $|\psi_2\rangle$ of the particles satisfy
\begin{eqnarray}
\langle p/m_1|\psi_1\rangle = \langle p/m_2|\psi_2\rangle,   \label{equivv}
\end{eqnarray}
i.e., their wave functions in the `velocity basis' coincide.
 Thus, we are led to an equivalent but stronger statement of the equivalence principle for quantum systems:
\\ \\
{\bf Equivalence principle for quantum systems, Version B:} Any two particles with the same velocity wave-function behave identically in free fall, irrespective of their masses.
 \\ \\
 By "behave identically" we mean that for any function $f(x,v)$ on the velocity phase space the expectation values $\langle f\rangle = \int dx dv f(x,v) \bar{W}(x,v)$ are the same irrespective of the mass.  In order to examine what this implies for the corresponding operators, let us denote by  ${\cal H}_m$ the Hilbert space associated to a particle of mass $m$ in the Schr\"odinger representation, i.e., ${\cal H}_m = L^2({\pmb R}, dx)$. Functions $f(x)$ on the phase space that do not depend on velocity are mapped to the same multiplicative operator $f(\hat{x})$ on ${\cal H}_m$. Thus, version B implies the same probabilities and expectation values for position measurements. However, a general function $f(x, v)$ is mapped to different operators
   $\hat{F}_1$ and $\hat{F}_2$ on the Hilbert spaces ${\cal H}_{m_1}$ and ${\cal H}_{m_2}$. The version B of the EP then implies that $\langle \psi|\hat{F}_1|\psi\rangle_{{\cal H}_{m_1}} =\langle \psi|\hat{F}_2|\psi\rangle_{{\cal H}_{m_2}}$.

Version B of the EP is logically distinct from Version A, even though the two versions coincide for the particular class of systems that is considered in this paper.  The two versions would differ, for example, for  free-falling particles in the presence of gravity-induced decoherence or dynamics that involve mass-spin coupling \cite{Lammer}. We think that Version B appears more intuitive than Version A, and it is probably easier to generalize to more complex set-ups that involve a full relativistic treatment of particles and strong gravitational fields.  Such a generalization will not be straightforward, though. The evolution of the Wigner function will not be as simple as (\ref{wigev}), because of the non-linear relation between velocity and momentum in relativistic systems.

Version B of the EP is operationally distinct from Version A, so that it can be tested in a different type of experiments. Version B involves the comparison of two experiments carried out on particles of different mass. Thus, it is closer to the classical statement of the EP. However, there is a caveat. Version B requires that we are able to prepare two different  types of particle in the same velocity wave function. It is not obvious how this can be effected in the general case. It is perhaps feasible in the special case where the two particles
 are composite, they have identical constitution and their mass difference can be attributed solely to the excitation of the states associated with the internal degrees of freedom.

There is no    direct correspondence between Hilbert spaces associated to particles of different mass. The reason is that particles of different mass correspond to unitarily inequivalent representations of the Galileo group (or the Poincar\'e group), and thus, there is no natural identification of observables defined on the different Hilbert spaces.

Given the restriction above, the choice of the velocity basis provides the most natural way of identifying states for particles of different mass, as long as relativity is fully taken into account. The reason is that the defining representation for a  relativistic particle of mass $m$ consists of square integrable functions on the positive-energy cone $V^+_m = \{ p_{\mu}, p_{\mu}p^{\mu} = m^2, p_0 >0 \}$. Since $V^+_m \cap V^+_{m'} = \emptyset$ for  $m \neq m'$, the only natural way of comparing the quantum states of particles with different mass is using wave functions defined with respect to the four-velocity $v^{\mu} = p^{\mu}/m$, since the four-velocity is a unit four-vector for all masses.


\section{A free-falling composite particle}

\subsection{The Hamiltonian}
 Next, we  consider the free fall of composite particles, i.e.,  particles with constituent elements, like atoms or  molecules. A composite particle has internal degrees of freedom in addition to the translational ones (position and momentum). The simplest model for the internal degrees of freedom is the two-level atom, which is  quintessential   for many atom-optical phenomena.

    We begin with the description of composite particles in QFT and then   come down to the present nonrelativistic weak gravity case. In this procedure, we avoid notions that are not essential to the quantum treatment of composite particles, like, for example the notion of the center of mass (which is non unique in relativistic systems),  or the definition of position and momentum operators for the constituent particles.

At the fundamental level, composite particles are described by interacting QFTs. In QFT, an observable particle of mass $m$ is conventionally defined as being associated with poles of the S-matrix at energy $E=mc^2$ in the rest frame of the system  \cite{PeSch}. If the pole is at a real energy, the mass is real and the particle is stable; if the pole is at a complex energy the mass is complex and the particle is unstable.

Consider the $S$-matrix restricted in a subspace with fixed conserved quantities like baryon number, lepton number,  or quantities that are approximately conserved at lower energies like number of nuclei of a given type. Such a subspace conveys the classical concept of a composite particle---the constituents are specified by the conserved quantities---with internal degrees of freedom. The S-matrix
 has
 a sequence of poles labeled by the integers $n = 0, 1, 2, \ldots$. Each pole is characterized by a different rest mass $m_n$, so that $m_0 < m_1 \leq m_2 \leq \ldots$, and by different values of spin $s_n$. A single composite particle for this QFT is described by the Hilbert space
\begin{eqnarray}
{\cal H} = \oplus_n {\cal H}_{m_n, s_n},
 \end{eqnarray}
 where ${\cal H}_{m, s}$ is the Hilbert space associated with an irreducible representation of the Poincar\'e group with mass $m$ and spin $s$.  The Hilbert space ${\cal H}_{m,s}$ can be written as ${\cal H}_0 \otimes {\pmb C}^{2s+1}$ where ${\cal H}_0 = L^2(V_1^+)$ contains square integrable wave functions over four-velocities $u^{\mu}$ with $u_0 >0$. Then,
 \begin{eqnarray}
 {\cal H} =  {\cal H}_0 \otimes {\cal H}_{int}, \label{hfac}
  \end{eqnarray}
  where ${\cal H}_{int} = \oplus_n C^{2s_n+1}$ describes all internal (i.e., non translational) degrees of freedom of the composite particles. The Hilbert space ${\cal H}_{int}$ is spanned by a basis $|n \rangle$, that defines the Hamiltonian for the internal degrees of freedom, $\hat{H}_{int} = \sum_n {m_n}|n\rangle \langle n|$.

The above applies to a   composite particle that moves freely. Next, we consider such a particle in a homogeneous gravitational field. We assume that the gravitational field is so weak that it does not change the internal states of the particle. Then, the Hamiltonian on each subspace ${\cal H}_{m_n}$ is given by
 Eq. (\ref{H1}) in the Appendix (or its analogue for particles with spin).  In the non-relativistic limit and restricting to one spatial dimension, Eq. (\ref{H1}) reduces to Eq. (\ref{Hamiltonian1}). Hence, we can express the Hamiltonian for a composite particle in a weak homogeneous gravitational field as a matrix with respect to the
  basis $|n\rangle$ of ${\cal H}_{int}$

\begin{eqnarray}
    \!\!\!\ \!\!\!\!\!\! \!\!\!\ \!\!\!\!\!\!   \hat{H}_g =  \left( \begin{array}{cccc} m_0 + \frac{\hat{p}^2}{2m_0} + m_0g \hat{x} &0 & 0& \ldots \\0&m_1 + \frac{\hat{p}^2}{2m_1} + m_1g \hat{x}&0 &\ldots \\ 0 & 0 &  m_2 + \frac{\hat{p}^2}{2m_2} + m_2g  \hat{x}&\ldots\\ \ldots &\ldots &\ldots & \ldots\end{array}\right).   \label{hammm}
\end{eqnarray}

Given Eq. (\ref{hfac}), a  general initial state $|\Psi_0\rangle $ is decomposed with respect to the basis $|n\rangle$ of ${\cal H}_{int}$ as

\begin{eqnarray}
|\Psi_0\rangle = \sum_n c_n |\psi_{n,0}\rangle \otimes |n\rangle \label{Psi0}
\end{eqnarray}
where the vectors $|\psi_{n, 0} \rangle \in {\cal H}_{0}$ correspond to the translational degrees of freedom.

The state $|\Psi_0\rangle$  evolves under the Hamiltonian (\ref{hammm}) to
\begin{eqnarray}
|\Psi_t^{(g)}\rangle = \sum_n c_n e^{i\frac{m_ng^2t^3}{3}} \hat{V}(-m_ngt, -\frac{1}{2}gt^2) |\psi_{n,t}^{(0)}\rangle \otimes |n\rangle, \label{timev}
\end{eqnarray}
where  $|\psi_{n,t}^{(0)}\rangle$ is the evolution of the initial state $|\psi_{n,0}\rangle$ with the free-particle Hamiltonian $\hat{H}_n = m_n + \frac{\hat{p}^2}{2m_n}$.

\subsection{Validity of the equivalence principle for composite particles}

Next, we consider measurements only of the translational degrees of freedom. All information about such measurement is encoded in the reduced density matrix on ${\cal H}_0$ that is obtained by a partial trace of the internal degrees of freedom

\begin{eqnarray}
\langle x|\hat{\rho}^{(g)}_{red}(t)|x'\rangle :=  \sum_n \langle x, n|\Psi_t^{(g)}\rangle \langle \Psi_t^{(g)}|x', n \rangle \nonumber \\
 = \sum_n |c_n|^2 e^{-im_ngt(x-x')} \psi_{n,t}^{(0)}(x+\frac{1}{2}gt^2) \psi_{n,t}^{*(0)}(x'+\frac{1}{2}gt^2).  \label{rhored}
\end{eqnarray}
The probability density for position is obtained from the diagonal elements of the reduced density matrix
\begin{eqnarray}
\hspace{-0.5cm} \langle x|\hat{\rho}^{(g)}_{red}(t)|x\rangle = \sum_n |c_n|^2 |\psi_{n,t}^{(0)}(x+\frac{1}{2}gt^2)|^2  = \langle x + \frac{1}{2} gt^2|\hat{\rho}^{(0)}_{red}(t)| x + \frac{1}{2} gt^2 \rangle. \label{posden}
\end{eqnarray}
  Eq. (\ref{posden}) manifestly satisfies Version A of the EP.

 Regarding Version B of the EP, we have to employ the velocity density matrix for the translational degrees of freedom. This is naturally defined, because of the splitting (\ref{hfac}) of the Hilbert space ${\cal H}$, since ${\cal H}_0$ is naturally defined in the velocity basis.

 It is simpler to work with the velocity Wigner function of composite particles. This is defined as follows.
     Let $|\Psi \rangle = \sum_n c_n |\psi_n\rangle \otimes |n\rangle$ be a general state on ${\cal H}$ and  let $\bar{W}_n(x,v)$ be the velocity Wigner function associated with the vectors  $|\psi_n\rangle$ according to Eq. (\ref{velwig}). Then, the reduced velocity Wigner function for the translational degrees of freedom is defined as
\begin{eqnarray}
\bar{W}_{red}(x,v) = \sum_n |c_n|^2 \bar{W}_n(x, v)
\end{eqnarray}

 We readily verify that the time evolution of $\bar{W}_{red}$ is given by Eq. (\ref{wigev2}). Thus,  Version B of the equivalence principle is also satisfied when expressed in terms of $\bar{W}_{red}$.

\subsection{Dephasing of the translational dof by the internal dof}

\subsubsection{The evolution of a factorized initial state}

Consider now the special case of a factorized initial state
\begin{eqnarray}
|\Psi_0\rangle = |\psi_0\rangle \otimes \sum_n c_n |n\rangle,  \label{factoriz}
 \end{eqnarray}
 i.e., a state where all vectors $|\psi_{n,0}\rangle$ in Eq. (\ref{Psi0}) coincide with $|\psi_0\rangle$.

Time evolution entangles the translational and internal degrees of freedom
\begin{eqnarray}
|\Psi_t^{(g)}\rangle = \sum_n e^{i\frac{m_ng^2t^3}{3}} \hat{V}(-m_ngt, -\frac{1}{2}gt^2) |\psi_{n,t}^{(0)}\rangle \otimes |n\rangle . \label{timev2}
\end{eqnarray}
In this case, the dependence of $|\psi_{n,t}^{(0)}\rangle$ on $n$ in Eq. (\ref{timev2}) is not due to the initial condition, but due to the fact that the time-evolution of any state is mass-dependent, and the mass depends on $n$. A measure of the dependence of $|\psi_{n,t}^{(0)}\rangle$ on $n$ is the difference $\delta_n$ in the position dispersion $\Delta x^2(t)$ between $|\psi_{n,t}^{(0)}\rangle$ and $|\psi_{0,t}^{(0)}\rangle$. By Eq. (\ref{dispers}),
\begin{eqnarray}
\delta_n^2 = \frac{t^2}{4 (\Delta x)_0^2} (m_0^{-2} - m_n^{-2}). \label{deltan}
\end{eqnarray}

We assume that the excitation energies
\begin{eqnarray}
\omega_n = m_n - m_0,
\end{eqnarray}
 are much smaller than the energy $m_0$ of the ground state. Then, Eq. (\ref{deltan}) becomes
 \begin{eqnarray}
 \delta_n^2 = \frac{t^2 \omega_n}{2m_0^3 (\Delta x)_0^2} .
 \end{eqnarray}

Observe that the states $|\psi_{n,t}^{(0)}\rangle$ are almost identical if $\delta_n << (\Delta x)_0$, i.e., for times
\begin{eqnarray}
t << \sqrt{m_0/\omega_n} m_0 (\Delta x)_0^2. \label{condimp}
\end{eqnarray}
 In this regime, all   $\psi_{n,t}^{(0)}(x)$ in Eq. (\ref{timev2})
  coincide with
\begin{eqnarray}
\psi_t^{(0)}(x) = \int dx' G_t^{(0)} (x,x') \psi_0(x'),
\end{eqnarray}
where the propagator $G^{(0)}$ is defined with respect to the mass $m_0$.

The state (\ref{timev2}) still remains entangled. Therefore, the reduced density matrix for the translational degrees of freedom is a mixed state,
\begin{eqnarray}
\langle x|\hat{\rho}^{(g)}_{red}(t)|x'\rangle :=
 \Gamma_t( x-x') e^{-im_0 gt(x-x')} \psi_t^{(0)}(x+\frac{1}{2}gt^2)\psi_t^{*(0)}(x'+\frac{1}{2}gt^2) \label{reddx},
\end{eqnarray}
where
\begin{eqnarray}
\Gamma_t(\Delta x) = \sum_n |c_n|^2 e^{-i \omega_n t g \Delta x}.
\end{eqnarray}

\subsubsection{Dephasing due to internal degrees of freedom}

Assume that the internal degrees of freedom are in a thermal state at temperature $\beta^{-1}$, whence $|c_n|^2 \sim  e^{-\beta \omega_n}$. Then,
\begin{eqnarray}
\Gamma_t(\Delta x)  = \frac{Z(\beta + i g t \Delta x)}{Z(\beta)},
\end{eqnarray}
where $Z(\beta) = \sum_n e^{-\beta \omega_n}$ is the partition function for the internal degrees of freedom.

For $g \beta^{-1} t \Delta x << 1$, we expand the partition function $\log Z(\beta + \delta) = \log Z(\beta) - \langle E\rangle \delta +\frac{1}{2} \beta^{-2} C_{\upsilon} \delta^2$, in terms of the mean energy $\langle E\rangle$ and the heat capacity $C_{\upsilon}$, to obtain
\begin{eqnarray}
|\Gamma_t(\Delta x)| \simeq e^{-\frac{1}{2} C_{\upsilon} (g \beta^{-1}(\Delta x) t)^2}.
\end{eqnarray}

Hence, time evolution typically suppresses the off-diagonal elements of the density matrix (\ref{reddx}), i.e., superpositions of states with position localizations that differ by $\Delta x$. The relevant time scale $\tau_d$ is
\begin{eqnarray}
\tau_d = \frac{\beta}{g \Delta x \sqrt{C_{\upsilon}}}.
\end{eqnarray}
There is no suppression at low temperatures, since $Z(\beta) \rightarrow 1$ as $\beta \rightarrow \infty$.


\subsubsection{Remarks}

Eq (34) describes a dephasing of the translational dof by the internal dof of the composite particle. It has the same form as the result obtained by Pikovsky et al \cite{PikBru15}. However,   we give a different physical interpretation, as a dephasing rather than a decoherence process---we clarify the difference below. This dephasing process is universal, in the sense that it applies to all particles and the relevant time-scale $\tau_d$ does not depend directly on the particle mass \cite{PikBru17} (only indirectly through the heat capacity of the internal degrees of freedom). For further discussion of this issue and critique, see Refs.  \cite{Zeh, GooUnr, AdlBas, Bonder, Diosi15}. Our derivation  is similar  to the analysis  of \cite{Pang16}, in that we did not employ the concept of gravitational time dilation.

We note that
  Eq. (\ref{reddx}) applies only to factorized initial states, and not to general states that are described by Eq. (\ref{rhored}). This means that the dephasing is strictly derived only for such states.   Since the Hamiltonian involves coupling between translational and internal degrees of freedom, the generic state for a composite particle involves some entanglement.  Nonetheless, continuity suggests that some degree of dephasing is present for a larger class of initial states, but its extent remains to be quantified. We also note that
  Eq. (\ref{reddx}) applies only as long as Eq. (\ref{condimp}) is satisfied. For sufficiently large times,  Eq. (\ref{condimp}) is violated and Eq. (\ref{timev2}) applies. These means that some degree of phase coherence might be restored at later times due to the fact that wave-function components with different masses manifest different dispersion.

Regarding the physical essence of this process, we believe that it is
  closer to  the dephasing in spin-echo experiments than to environment-induced decoherence.
  In spin-echo experiments, the phase coherence of a system with many   degrees of freedom is lost because of the inhomogeneity of an external magnetic field that acts upon the particle spins. However, this process involves no information loss: information is stored in the correlations between spin and particle position \cite{AnSav11}. A suitable manipulation of the external field can restore the phase coherence. In the limit where dissipation can be ignored, information (and phase coherence) is fully restored.

   To see the analogy with the present case, assume that an initial state of the form (\ref{factoriz}) evolves to time $T$ within a homogeneous gravitational field $g$.  Let $T$ be sufficiently small so that     Eq. (\ref{condimp}) is satisfied. Then, the state  vector at time $T$ in the position basis is
  \begin{eqnarray}
  \sum_n c_n e^{-im_ngTx - \frac{im_ng^2T^3}{6}} \psi_T^{(0)}(x+ \frac{1}{2}gT^2)\otimes |n\rangle. \nonumber
  \end{eqnarray}
  Suppose that at time $t = T$, the gravitational field is inverted and the system evolves to time $t = 2 T$. The state vector at $t = 2 T$ is
  \begin{eqnarray}
  \psi_{2T}^{(0)} (x)\otimes \sum_n c_n e^{-2im_n T - i\frac{ m_ng^2T^3}{3}}|n\rangle, \nonumber
   \end{eqnarray}
   where $\psi_{2T}^{(0)} (x)$ stands for the initial state evolved to time $2T$ in absence of a gravitational field. Thus,
   all phase information will have been restored to the translational degrees of freedom. If the initial state is a cat state, it will appear to be  `decohered' at $t = T$, but then the full cat will reappear  at $t = 2T$, i.e., we will observe `recoherence'.

One may object that  it is impossible to actually invert the gravitational field, either by natural processes or in the laboratory. In reply we note that  recoherence only requires  that the effective `force' on the particle is inverted.  This can be achieved, for example, by assuming that the particle has charge $q$, and that at time $T$ a homogenous electric field ${\cal E}$ is switched on. If $q {\cal E} = 2 m g$, the change to the potential is equivalent to an inversion of $g$.

The key point is that---as in the spin-echo case---the loss of phase coherence is not accompanied by loss of information. The information persists in the correlations between the translational and the internal degrees of freedom, and it can be recovered using only local operations upon the particle. This behavior contrasts   most models of environment-induced decoherence, where the lost information is maximally distributed over the environmental degrees of freedom \cite{GeHa, Hal, DHal} and it cannot be fully restored with local operations.


\section{A free-falling atom: Qubit phase shift}

\subsection{Derivation}
In the previous section, we focused on measurements on the translational degrees of freedom. Here, we focus on the internal degrees of freedom. For simplicity, we consider a composite particle with only two internal states, representing, for example, an atomic qubit.

Consider a factorized initial state (\ref{factoriz}), and assume that Eq. (\ref{condimp}) applies. We can evaluate the reduced density matrix for the internal degrees of freedom
\begin{eqnarray}
\hat{\rho}_{qub}(t) = \left( \begin{array}{cc} |c_0|^2 & c_0 c_1^* e^{i\omega t - i\omega g^2 t^3/3} \zeta_t \\
c_0^*c_1 e^{-i \omega t + i\omega g^2 t^3/3} \zeta_t^* & |c_1|^2\end{array},
 \right),  \label{rhoqub}
\end{eqnarray}
where
\begin{eqnarray}
\zeta_t = \int dx |\psi_t^{(0)}(x)|^2 e^{-i\omega gtx}, \label{zetat}
\end{eqnarray}
is a function of time which encapsulates the degree of `quantumness' of the initial quantum state.

 As a first example, we consider a  initial state that is essentially classical:  it is localized around $x = 0$, with position spread $\sigma_x$ and with vanishing mean momentum. Assume that the qubit is recorded at distance $L$ from each source. Hence, the detection time is strongly peaked around $t_d = \sqrt{2L/g}$. If
 \begin{eqnarray}
b := \omega g \sqrt{2L/g} \sigma_x << 1, \label{cond0}
 \end{eqnarray}
then $\zeta_{t_d} \simeq 1$, and it can be ignored in Eq. (\ref{rhoqub}).

Hence, the qubit density matrix has developed a phase due to the free fall,
\begin{eqnarray}
\phi_g = \frac{1}{3} \omega g^2 t_d^3 = \frac{2\sqrt{2}}{3} \omega g^{1/2} L^{3/2}, \label{phig}
\end{eqnarray}
 in addition to the phase $\omega t$ due to free evolution.

This phase $\phi_g $ is  measurable, at least in principle. For $\omega$ in the microwave range, and $L$ of the order of $100 m$, $\phi_g$ varies between $10^{-4}$ and $10^{-2}$ radians. If $\phi_g$ is of order unity or smaller,   the condition (\ref{cond0}) is always satisfied since $b/\phi_g \sim \sigma_x / L << 1$. Thus, we predict a rotation of a qubit's Bloch vector by a phase $\phi_g$ due to free fall. The relative size of the phase shift
\begin{eqnarray}
u = \frac{\phi_g}{\omega t} =\frac{1}{3} g^2 t_d^2 = \frac{2}{3} gL
\end{eqnarray}
is frequency independent.

For $L = 100m$, $u = 7 \cdot 10^{-15}$, which is  well above the relative accuracy $\sim 10^{-16}$ of atomic fountain clocks \cite{NIST}. However, the atomic  fountain set-up cannot be itself employed for measurement of this effect. In atomic fountains,  $L \sim 1m$. This leads to $u = 7 \cdot 10^{-17}$, which is significantly below present accuracy.

We emphasize that the phase shift $\phi_g$    applies to all particles (irrespective of the nature of the forces between the constituents) and it does not depend on the mass $m_0$ of the composite system. Hence, the ratio $\phi_g/\omega$ will be the same for all atoms falling from the same height $L$.

\subsection{Classical relativistic analogue of the phase shift}
Expressing the phase shift $\phi_g$ in SI units,
\begin{eqnarray}
\phi_g =  \frac{2\sqrt{2}}{3} \frac{\omega g^{1/2}L^{3/2}}{c^2},
\end{eqnarray}
we note that it does not depend explicitly on $\hbar$. Thus, its origin is essentially classical. Indeed, it has a classical interpretation, namely,  half of $\phi_g$  originates from gravitational red-shift and half from special-relativistic time dilation \cite{Mu10}.

 To see this, consider the radial free-fall  of a particle in Schwarzschild spacetime which models the gravitational field of the Earth. The proper time $\tau$ of the particle is related to the coordinate time $t$ and the radial coordinate $r$ by
\begin{eqnarray}
d\tau^2 = (1-\frac{2GM}{r} )dt^2 - \frac{dr^2}{1-\frac{2GM}{r}}. \label{freef}
\end{eqnarray}
We rewrite Eq. (\ref{freef}) as
\begin{eqnarray}
d \tau = dt \sqrt{1 - \frac{2GM}{r} - \frac{v^2}{1-\frac{2GM}{r}}}, \label{proptime}
\end{eqnarray}
where  $v = dr/dt$. For a weak gravitational field ($GM/r<<1$) and non-relativistic velocity ($v<< 1$), we expand the square root in Eq. (\ref{proptime}) to obtain
\begin{eqnarray}
d \tau \simeq dt (1 - \frac{GM}{r} - \frac{1}{2}v^2).
\end{eqnarray}

Suppose we drop a body from   $r = R$. For  $r = R - x$ with $x << R$,
\begin{eqnarray}
d \tau \simeq dt( 1 - \frac{GM}{R} - g x - \frac{1}{2}v^2),
\end{eqnarray}
where $g = GM/R^2$ is the gravitational acceleration, approximately constant as long as $x << R$.

Let the trajectory of the falling particle be given by the function $x(t)$. Then,
\begin{eqnarray}
\tau = (1 - \frac{GM}{R}) t - g \int_0^t ds x(s) - \frac{1}{2} \int_0^t ds \dot{x}^2(s) \label{prop2}
\end{eqnarray}
The term $g \int_0^t ds x(s)$ corresponds to gravitational time-dilation and the term $\frac{1}{2} \int_0^t ds \dot{x}^2(s)$ to  special-relativistic time dilation. For a general path $x(s)$ those two terms are different. However, for a free-falling particle, $x(t) = \frac{1}{2}gt^2$, both terms turn out to be equal to $\frac{1}{6}g^2t^3$, so that
\begin{eqnarray}
\tau =  (1 - \frac{GM}{R}) t  - \frac{1}{3}g^2t^3.
\end{eqnarray}
Thus, the phase shift for a qubit of frequency $\omega$ in the rest frame is
\begin{eqnarray}
\omega \tau = \omega (1 - \frac{GM}{R}) t - \phi_g = \omega \tau_0 - \phi_g, \label{clphs}
\end{eqnarray}
where the phase shift $\phi_g = \frac{1}{3}\omega g^2t^3$ coincides with that of Eq. (\ref{phig}) and $\tau_0$ is the proper time for a static observer at  $r = R$.

\subsection{Phase shift for non-classical states}
The equivalence between the phase shift of Eq. (\ref{rhoqub}) and the classical phase shift (\ref{clphs}) holds only for quantum states that behave classically. Such states have  Wigner functions that are well localized in phase space, so that their time evolution has a simple correspondence to classical paths. However,
 Eq. (\ref{rhoqub}) is more general as it also applies to non-classical states.

As an example, we consider an initial state $\psi_0(x) = \frac{1}{\sqrt{2}}[\psi_{10}(x) + \psi_{20}(x)]$, where $\psi_{i0}(x)$ are classical states, localized in phase space around positions $x_i$ and velocities $v_i$, for $i = 1, 2$. It is convenient to choose a reference frame so that $x_1 = -x_2 = \frac{1}{2} \ell$. We assume that the positions spread $\sigma_x$ of each component is sufficiently small that the two states $\psi_{i0}(x)$ do not overlap. Then, the i-th component of the  wave function corresponds to a path $x_i(t) = x_i + v_i t$ under free evolution. For  $t << g \sigma_x \omega t $, the parameter $\zeta_t$ of Eq. (\ref{zetat}) is
\begin{eqnarray}
\zeta_t = e^{-\frac{i}{2} g\omega (v_1+v_2) t^2} \cos\left[\frac{1}{2}g \omega t [\ell + (v_1-v_2)t] \right]. \label{zetat2}
\end{eqnarray}
We consider two special cases. The first case corresponds to $v_1 = v_2 = 0$, i.e., we have no momentum superpositions. Then the phase shift $\phi_g$ remains the same, but the norm of the off-diagonal elements oscillates
 \begin{eqnarray}
 |\rho_{01}(t)| = |\rho_{01}(0)| \left|\cos\left(\frac{1}{2}g \omega  \ell t \right)\right|.
 \end{eqnarray}
 The second case corresponds to a measurement at time $t_d = \ell/(v_2-v_1)$, so that both components of the wave function arrive simultaneously at a detector located at  $x = -L$, where $L = \frac{1}{2}gt_d^2 - \frac{1}{2} \frac{v_2+v_1}{v_2-v_1} \ell$. Then, the cosine in Eq. (\ref{zetat2}) becomes unity, and the off-diagonal terms exhibit a phase shift
 \begin{eqnarray}
 \phi_g = \frac{1}{3}\omega g^2 t_d^3 + 2 \omega L \frac{v_1+v_2}{2}   + \frac{1}{2} g\omega \ell \frac{(v_1+v_2)^2}{v_2-v_1}.
 \end{eqnarray}
 The first term is the previously derived phase shift. The second term is a classical, non-gravitational contribution that originates from the non-zero mean velocity of the initial state. The third term is genuinely quantum as it incorporates the contribution from the non-classical character of the initial state. Again, $\phi_g$ does not depend on the mass of the falling particle.

\section{Conclusions}
The key findings of this paper  are the following.

First, we come up with two statements of the EP for quantum systems:\\
Version \textbf{A}: The probability distribution of position for a free-falling particle is the same as the probability distribution of a free particle, modulo a {\em mass-independent} shift of its mean.\\
Version \textbf{B}: Any two particles with the same velocity wave-function behave identically in free fall, irrespective of their masses.

Both statements involve the comparison between the outcomes of different experiments. They are fully equivalent at the level of elementary calculations. However, they are logically and operationally distinct. They can be tested in different experimental set-ups and they suggest different generalizations.

Second, we studied the effect of free fall on the particle's internal degrees of freedom. Assuming only mass-energy equivalence, free fall induces a coupling between internal and translational degrees of freedom. The implications of such a coupling depend on the initial state of the system and on the observable that is being measured.

 For a particular class of initial states, we verify that the internal degrees of freedom can lead to suppression of off-diagonal terms of the density matrix in the position basis.   We argue that this phenomenon   is in the nature of dephasing and not environment-induced decoherence, because it does not involve irreversible loss of information.

Third, we found a gravitational phase shift in the reduced density matrix of the internal degrees of freedom. While this phase shift is  a fully quantum effect, it has a natural classical interpretation in terms of gravitational red-shift and special relativistic time-dilation. However, it is also defined for states with no classical analogue.

\vspace{.5cm}

{\bf Acknowledgment}  We thank   Albert Roura for helpful discussions, in particular, about the measurability of the phase shift. We also thank
 Igor Pikovsky and Magdalena Zych for explaining their work. C. A. acknowledges support by Grant No. E611 from the Research Committee of the University of Patras via the ”K. Karatheodoris” program.

\begin{appendix}

\section{The single-particle Hamiltonian}

 We show in this Appendix how Eq. (1) is derived from quantum field theory, which serves two additional purposes: First, we show that QFT does not generate anything new or different in the nonrelativistic weak field limit, from the known quantum mechanical description. No new quantum equivalence principle is hidden in a more sophisticated field theory description.  Second, the weak equivalence principle formulated in classical physics is assumed to be valid and used in the Lagrangian of the scalar field, as shown below.

Consider a static spacetime
\begin{eqnarray}
ds^2 = - N^2(x) dt^2 + h_{ij}(x) dx^i dx^j, \label{metric}
\end{eqnarray}
where $N$ is the lapse function and $h_{ij}$ is the three-metric on a spacelike surface $\Sigma$.

We study a minimally coupled scalar field with Lagrangian density
\begin{eqnarray}
{\cal L} = \sqrt{-g} (- \frac{1}{2} g^{\mu \nu} \nabla_{\mu}\phi \nabla_{\nu}\phi -\frac{1}{2}m^2 \phi^2)
\end{eqnarray}
on the metric (\ref{metric}).
Since the spacetime is static, there is a unique quantization with respect to the Killing field $\frac{\partial}{\partial t}$.


To quantize, we first construct the classical  Hamiltonian
\begin{eqnarray}
H = \frac{1}{2} \int d^3 x N\sqrt{h} \left(p^2 +h^{ij}\partial_i\phi \partial_j\phi +m^2\phi^2\right), \label{hcl}
\end{eqnarray}
where $p$ is the conjugate momentum of $\phi$ that satisfies
\begin{eqnarray}
\{ \phi(x), p(x')\}  = \delta^3(x, x').
\end{eqnarray}
We define the inner product
\begin{eqnarray}
(\psi, \phi) = \int d^3x \sqrt{h} \psi^*(x) \phi(x).
\end{eqnarray}
 on the square-integrable functions of $\Sigma$. The Hamiltonian (\ref{hcl}) becomes
 \begin{eqnarray}
 H = \frac{1}{2} (p,Np) + \frac{1}{2} (\phi, K\phi),
 \end{eqnarray}
 where
 \begin{eqnarray}
 K\phi = -Ν \nabla^2 \phi - \partial_iN h^{ij}\partial_j\phi + N m^2 \phi.
 \end{eqnarray}

 To quantize, we express $\phi$ and $p$ as operators on a Fock space, defined in terms of the creation and  annihilation operators
 \begin{eqnarray}
 \hat{a}(x) = \frac{1}{\sqrt{2}} H_1^{-1/2}\left(K^{1/2} \hat{\phi} - iN^{1/2} \hat{p}\right) \\
 \hat{a}^{\dagger}(x) = \frac{1}{\sqrt{2}} H_1^{-1/2}\left(K^{1/2} \hat{\phi} + iN^{1/2} \hat{p}\right),
 \end{eqnarray}
 where
 \begin{eqnarray}
 H_1 = \frac{1}{2} (\sqrt{K}\sqrt{N} +\sqrt{N} \sqrt{K}). \label{H1}
 \end{eqnarray}
 Then the quantized field Hamiltonian becomes
 \begin{eqnarray}
 \hat{H} = \int d^3x \sqrt{h} \hat{a}^{\dagger}(x) H_1\hat{a}(x).
 \end{eqnarray}
 Thus, $H_1$ is the Hamiltonian for a single particle.

 We consider the Newtonian gravity limit, in which $h_{ij} = \delta_{ij}$ and $N = 1 + U$, where $U << 1$ is the Newtonian potential. Then, $\hat{p}_i = -i \partial_i$, and the operator $\hat{K}$ becomes
 \begin{eqnarray}
 \hat{K} =  (1+U) (m^2 +\hat{\pmb p}^2) - i {\pmb \nabla U}  \cdot \hat{\pmb p}.
 \end{eqnarray}
 In the non-relativistic limit,
 \begin{eqnarray}
 \sqrt{K} = (1+U) [m + \frac{\hat{\pmb p}^2}{2m}] - \frac{i}{2m} {\pmb \nabla U}  \cdot \hat{\pmb p}
 \end{eqnarray}
By Eq. (\ref{H1}),  $\hat{H}_1 = \sqrt{K} + \frac{1}{2} m U$, so that
 \begin{eqnarray}
 \hat{H}_1 = m + \frac{\hat{\pmb p}^2}{2m} + m \hat{U} -  \frac{1}{2m} \hat{p}\hat{U}\hat{p} \label{H1b}
 \end{eqnarray}
 This is the expected expression for the non-relativistic Hamiltonian of a particle in a Newtonian gravitational field, together with the leading correction term.

The Hamiltonian (\ref{H1b}) is characterized by an equality between inertial and gravitational mass. In order to obtain a QFT where inertial and gravitational mass differ, we have to start with a field-gravity coupling that is incompatible with General Relativity. Consider for example the Lagrangian,
\begin{eqnarray}
{\cal L} =  - \frac{1}{2} \eta^{\mu \nu} \partial_{\mu}\phi \partial_{\nu}\phi -\frac{1}{2}m^2 \phi^2 +  \lambda U(x) \phi^2, \label{lagr}
\end{eqnarray}
where $\eta_{\mu \nu}$ is the Minkowski metric, $\lambda $ is the gravitational coupling constant (proportional to the gravitational constant $G$)  and $U(x)$ a scalar field that describes the field coupling to gravity. Eq. (\ref{lagr}) is Poincar\'e covariant, but not generally covariant. The corresponding QFT can be constructed using standard canonical quantization. The one-particle Hamiltonian in the non-relativistic limit
\begin{eqnarray}
 \hat{H}_1 = m + \frac{\hat{\pmb p}^2}{2m} + \frac{\lambda}{m} U(\hat{x}).
\end{eqnarray}
 Assuming that $U$ coincides with the gravitational potential, the inertial and the gravitational mass are in general unequal. More importantly, their ratio differs for different particles unless one postulates that $\lambda$ is universal, i.e., the same for all types of particle.

\end{appendix}

\vspace{1cm}


\begin{thebibliography}{999}

\bibitem{DaFa82}  P.C.W. Davies and  J. Fang, Quantum Theory and the Equivalence Principle. Proc. R. Soc. Lond.
A381, 469 (1982).

\bibitem{CanSci83} P. Candelas and D. Sciama, Is there a Quantum Equivalence Principle?, Phys. Rev. D8, 1715 (1983)







\bibitem{CaMa97} C. Alvarez and  R. Mann, Testing the Equivalence Principle in the Quantum Regime. Gen. Relativ.
Gravit. 29, 245 (1997).

\bibitem{DalMaz} D. A. R. Dalvit and F. D. Mazzitelli, Quantum Corrected Geodesics, Phys. Rev. D60, 084018 (1999).

\bibitem{SinWil} D. Singleton and S.  Wilburn, Hawking Radiation, Unruh Radiation, and the Equivalence Principle, Phys. Rev. Lett 107, 081102 (2011).


\bibitem{COW} R. Colella, A. W. Overhauser, and S. A. Werner, Observation of Gravitationally Induced Quantum Interference, Phys. Rev. Lett. 34, 1472 (1975).

\bibitem{PCC99}A. Peters, K. Y. Chung, and S. Chu, Measurement of Gravitational Acceleration by Dropping Atoms,
Nature 400, 849 (1999).


\bibitem{FDHW}  S. Fray, C. A. Diez, T. W. H\"ansch, and M. Weitz, Atomic Interferometer with Amplitude Gratings of Light and Its Applications to Atom Based Tests of the Equivalence Principle,
Phys. Rev. Lett. 93, 240404 (2004).

\bibitem{SHA}  D. Schlippert, J. Hartwig, H. Albers, L. Richardson, C. Schubert, A. Roura, W. Schleich, W.
Ertmer, E. Rasel, Quantum Test of the Universality of Free Fall. Phys. Rev. Lett. 112, 203002
(2014).





\bibitem{OnoVio} L. Viola  and R. Onofrio,  Testing the Equivalence Principle through Freely Falling Quantum Objects, Phys. Rev. D55, 455 (1997).

\bibitem{Dav04}  P.C. Davies, Quantum Mechanics and the Equivalence Principle. Class. Quantum Gravity 21,
2761 (2004).

\bibitem{Greenb} D.M. Greenberger, The Role of Equivalence in Quantum Mechanics. Ann. Phys. 47, 116
(1968).

\bibitem{ZycBru15}
M. Zych and C. Brukner, Quantum Formulation of the Einstein Equivalence Principle, arXiv:1502.00971.

\bibitem{Zych} M. Zych, Quantum Systems under Gravitational Time Dilation, (Springer, 2017).

\bibitem{OMMP}  P. J. Orlando, R. B. Mann, K. Modi. and F. A. Pollock, A Test of the Equivalence Principle(s) for Quantum Superpositions, Class. Quant. Grav.  33, 19LT01 (2016).

\bibitem{GeHa}  M. Gell-Mann and J.  B.  Hartle,     Classical Equations
for Quantum Systems,  Phys. Rev.   D47, 3345 (1993).

\bibitem{PikBru15}
I. Pikovski, M. Zych, F. Costa, and C. Brukner, Universal Decoherence due to Gravitational Time Dilation, Nat. Phys. 11, 668 (2015).


\bibitem{PikBru17} I. Pikovski, M. Zych, F. Costa, and C. Brukner, Time Dilation in Quantum Systems and Decoherence, New J. Phys. 19  025011 (2017).



\bibitem{Shresta} S. Shresta and B. L. Hu, Moving Atom-Field Interaction: Quantum Motional Decoherence and Relaxation, Phys. Rev. A 68 (2003) 012110

\bibitem{GBH} C.  R. Galley, R.  Behunin and B. L. Hu, Oscillator-Field Models of Moving Mirrors in Quantum Optomechanics,  Phys. Rev. A87, 043832 (2013).

\bibitem{SLH}  K. Sinha, S. Y. Lin and B. L. Hu, Mirror-Field Entanglement in a Microscopic model for Quantum Optomechanics,
Phys. Rev. A 92, 023852 (2015).


\bibitem{ML} J. C. Muga and J. R. Leavens, Arrival Time in Quantum Mechanics, Phys. Rep. 338, 353 (2000).

 \bibitem{ToA} J. C. Muga, R. S. Mayato, and I. L. Equisquiza, {\em Time in Quantum Mechanics, vol 1} (Springer 2008); J. G. Muga, A Ruschhaupt and A. Del Campo, {\em Time in Quantum Mechanics, vol 2} (Springer 2010).

\bibitem{AnSav} C. Anastopoulos and N. Savvidou,   Time-of-arrival Probabilities for General Particle Detectors, Phys. Rev. A86, 012111 (2012);
 Quantum Temporal Probabilities in Tunneling Systems, Ann. Phys. 336, 281 (2013);  Time-of-Arrival Correlations, Phys. Rev. A95, 032105 (2017).

\bibitem{OPM17} P.  J. Orlando, F. A. Pollock, and K. Modi,  How does Interference Fall?,	arXiv:1610.02141.

\bibitem{Karol}  F. Karolyhazy, Gravitation and Quantum Mechanics of Macroscopic Objects, Nuovo Cim. 52, 390 (1966); F.  Karolyhazy,  A.  Frenkel,  and  B.  Lukacs, On  the  Possible Role  of  Gravity  in
the Reduction  of the  Wave Function,  in Quantum Concepts in Space and Time, R. Penrose and C. J. Isham editors, (Oxford,
1986, Clarendon Press).

\bibitem{Diosi}  L. Diosi, Gravitation and Quantum-Mechanical Localization of Macro-Objects, Phys. Lett. A105, 199 (1984); A Universal Master Equation for the Gravitational Violation of Quantum Mechanics, Phys. Lett. 120, 377 (1987); Models for Universal Reduction of Macroscopic Quantum Fluctuations, Phys. Rev. A40, 1165 (1989).


\bibitem{Penrose}  R. Penrose, Gravity and State Vector Reduction,  in Quantum Concepts in Space and Time, R. Penrose and C. J. Isham editors, (Oxford,
1986, Clarendon Press); On Gravity's Role in Quantum State Reduction,  Gen. Rel. Grav. 28, 581 (1996).

\bibitem{Bush}P. Bush,  Unsharp Localization and Causality in Relativistic
Quantum Theory. J. Phys. A: Math. Gen. 32, 6535 (1999).

\bibitem{Lammer} C. L\"ammerzahl, Quantum Tests of the Foundations of General Relativity, Class. Quant. Grav.
15, 13 (1998).

\bibitem{PeSch} M. E. Peskin and D. V. Schroeder, An Introduction To Quantum Field Theory (Addison Wesley, 2015), pg. 236.

\bibitem{Zeh} H. D.  Zeh,    Comment on Decoherence by Time Dilation, arXiv:1510.02239.

\bibitem{GooUnr} C. Gooding and W. H. Unruh, Bootstrapping Time Dilation Decoherence, Found. Phys. 45, 1166 (2015).

\bibitem{AdlBas} S. L. Adler and A. Bassi, Gravitational Decoherence for Mesoscopic Systems, Phys. Lett. A 380, 390 (2016).


\bibitem{Bonder}  Y. Bonder, E.  Okon, and D.  Sudarsky, Can Gravity Account for the Emergence of Classicality? Phys. Rev. D 92, 124050 (2015).

\bibitem{Diosi15}L. Diosi, Centre of Mass Decoherence Due to Time Dilation: Paradoxical Frame-Dependence, arXiv:1507.05828.

\bibitem{Pang16}
B. H. Pang, Y.Chen, and F. Ya. Khalili, Universal Decoherence under Gravity: A Perspective through the Equivalence Principle, Phys. Rev. Lett. 117, 090401 (2016).


\bibitem{spinecho} E. L. Hahn, Spin Echoes, Phys. Rev. 80, 580 (1950).

\bibitem{AnSav11} C. Anastopoulos and N. Savvidou,   Consistent Thermodynamics for Spin Echoes, Phys. Rev. E83, 021118 (2011).





\bibitem{Hal} J. J. Halliwell, Somewhere in the Universe: Where is the Information Stored When Histories Decohere?, Phys. Rev. D 60, 105031 (1999).

\bibitem{DHal} P. J. Dodd and J. J. Halliwell, Decoherence and Records for the Case of a ScatteringEnvironment, Phys. Rev. D 67, 105018 (2003).

\bibitem{NIST} T. P. Heavner, E. A. Donley, F. Levi, G. Costanzo, T. E. Parker,  J. H. Shirley,  N. Ashby,
S. Barlow
and S. R. Jefferts, First Accuracy Evaluation of NIST-F2, Metrologia
51, 174 (2014).

\bibitem{Mu10} H.  M\"uller, A. Peters, and S. Chu, A Precision Measurement of the Gravitational Redshift by the Interference of Matter Waves, Nature 463, 926 (2010).

\end{thebibliography}
\end{document}